\documentclass[aps, prd, amsmath, floats, floatfix, twocolumn, nofootinbib, nosuperscriptaddress]{revtex4-1}
\usepackage[pdftex]{graphicx}
\usepackage{color}
\usepackage{latexsym}
\usepackage[colorlinks]{hyperref}

\newcommand{\be}{\begin{eqnarray}}
\newcommand{\ee}{\end{eqnarray}}

\begin{document}

\title{Testing General Relativity with black hole X-ray data: a progress report}

\author{Cosimo~Bambi}
\email[]{bambi@fudan.edu.cn}
\affiliation{Center for Field Theory and Particle Physics and Department of Physics, Fudan University, 200438 Shanghai, China}

\begin{abstract}
Einstein's theory of General Relativity is one of the pillars of modern physics. For decades, the theory has been mainly tested in the weak field regime with experiments in the Solar System and observations of binary pulsars. Thanks to a new generation of observational facilities, the past 5~years have seen remarkable changes in this field and there are now numerous efforts for testing General Relativity in the strong field regime with black holes and neutron stars using different techniques. Here I will review the work of my group at Fudan University devoted to test General Relativity with black hole X-ray data.
\end{abstract}

\maketitle


\section{Introduction}

Einstein's theory of General Relativity was proposed at the end of 1915~\cite{Einstein:1916vd}. After more than 100~years and without any modification, General Relativity is still one of the pillars of modern physics. For decades, the theory has been mainly tested in the so-called weak field regime with experiments in the Solar System and observations of binary pulsars~\cite{Will:2014kxa}. In the past 20~years, there have been important efforts to verify the predictions of General Relativity on large scales with cosmological tests, mainly motivated by the problems of dark matter and dark energy~\cite{Jain:2010ka,Koyama:2015vza,Ferreira:2019xrr}. More recently, the interest has shifted to test General Relativity in the strong field regime with black holes and neutron stars~\cite{Bambi:2015kza,Yagi:2016jml}. The past 5~years have indeed seen significant changes in the field. Thanks to a new generation of observational facilities, we can now test General Relativity in the strong field regime with gravitational waves (see, e.g.,~\cite{TheLIGOScientific:2016src,Yunes:2016jcc,LIGOScientific:2019fpa}), mm VLBI data (see, e.g.,~\cite{Davoudiasl:2019nlo,Bambi:2019tjh,Psaltis:2020lvx}), and X-ray observations (see, e.g.,~\cite{Cao:2017kdq,Tripathi:2018lhx,Tripathi:2020dni}).

In General Relativity, black holes are simple objects and are completely characterized by three parameters, which are associated, respectively, to the mass, the spin angular momentum, and the electric charge of the object. This is the result of the no-hair theorem, which is actually a family of theorems, and holds under specific assumptions~\cite{Carter:1971zc,Robinson:1975bv,Chrusciel:2012jk}. The spacetime around an astrophysical black hole is thought to be described well by the Kerr solution~\cite{Kerr:1963ud}, where the compact object is specified by its mass $M$ and spin angular momentum $J$ while its electric charge vanishes. Indeed, the spacetime around a black hole formed from the gravitational collapse of some progenitor body is thought to quickly approach the Kerr solution by emitting gravitational waves. Deviations from the Kerr metric induced by a possible accretion disk or nearby stars are normally completely negligible for the spacetime geometry near the black hole event horizon. The equilibrium electric charge is normally negligible for a macroscopic object in a highly ionized host environment. Simple estimates of the deviations from the Kerr geometry around a black hole induced by these effects can be found, for instance, in Refs.~\cite{Bambi:2008hp,Bambi:2014koa,Bambi:2017khi}. On the other hand, macroscopic deviations from the Kerr metric are predicted by a number of scenarios involving new physics, from models with macroscopic quantum gravity effects at the black hole event horizon (see, e.g.,~\cite{Dvali:2012rt,Giddings:2014ova,Giddings:2017jts}) or exotic matter fields (see, e.g.,~\cite{Herdeiro:2014goa}) to scenarios of modified theories of gravity (see, e.g.,~\cite{Kleihaus:2011tg,Ayzenberg:2014aka}).

Here I will review the work of my group at Fudan University to use black hole X-ray data for testing General Relativity in the strong field regime. Our astrophysical systems are stellar-mass black holes in X-ray binaries or supermassive black holes in active galactic nuclei (AGN) accreting from geometrically thin and optically thick accretion disks.


\section{Testing black holes with X-ray data}

\begin{figure}[t]
\begin{center}
\includegraphics[width=8.5cm,trim={0cm 0cm 0cm 0cm},clip]{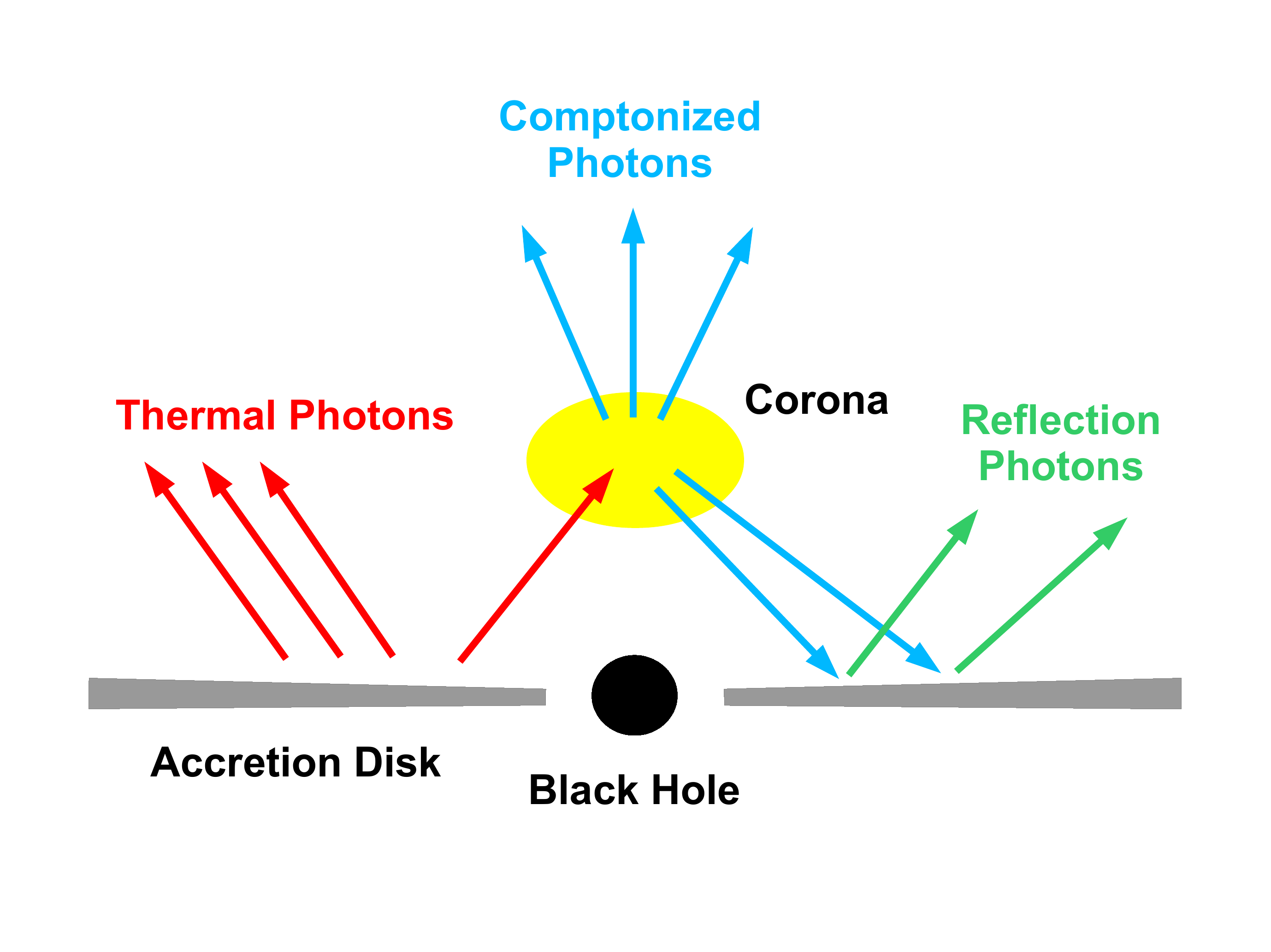}
\end{center}
\vspace{-1.2cm}
\caption{Astrophysical system for our tests. The black hole is accreting from a geometrically thin and optically thick cold accretion disk. A hot corona is near the black hole and the inner edge of the disk. The electromagnetic spectrum of the source is characterized by a thermal component from the disk, Comptonized photons from the corona, and a reflection spectrum from the disk. See the text for more details. \label{f-corona}}
\end{figure}

Fig.~\ref{f-corona} shows the astrophysical system for our tests (for more details, see~\cite{Bambi:2020jpe} and references therein). We have a black hole accreting from a geometrically thin and optically thick accretion disk. The gas in the disk is in local thermal equilibrium and every point on the surface of the disk emits a blackbody-like spectrum. The whole disk has a multi-temperature blackbody-like spectrum, which is peaked in the soft X-ray band (0.1-1~keV) for stellar-mass black holes and in the UV band (1-100~eV) for supermassive black holes. The ``corona'' is some hotter ($\sim$100~keV) plasma near the black hole. For example, it may be the atmosphere above the accretion disk, some gas in the plunging region between the black hole and the inner edge of the disk, the base of the jet, etc. A fraction of the thermal photons of the accretion disk inverse Compton scatter off free electron in the corona. The Comptonized photons illuminate the disk: Compton scattering and absorption followed by fluorescent emission generates the reflection spectrum of the disk.

The thermal and the reflection spectra of the accretion disk can be calculated within a theoretical model. Some calculations depend on the gravity theory and some assumptions valid in General Relativity may not hold in other theories of gravity. If we assume General Relativity, we have that:
\begin{enumerate}
\item the spacetime metric is described by the Kerr solution (Kerr black hole hypothesis);
\item particles follow the geodesics of the spacetime (Weak Equivalence Principle);
\item atomic physics in the strong gravitational field of the black hole is the same as that in our laboratories on Earth (Local Lorentz Invariance and Local Position Invariance).
\end{enumerate}
In the presence of new physics, one (or more) of these assumptions may be violated. We can thus think of constructing our theoretical model without relying on one (or more) of these assumptions. We can then compare the theoretical predictions of our new model with observational data and check whether the observations can confirm the assumptions valid in General Relativity. We can thus potentially test if the spacetime around an astrophysical black hole is described by the Kerr solution, if the particles of the gas in the accretion disk and the photons emitted by the accretion disk follow the geodesics of the Kerr spacetime, and if atomic physics near the black hole event horizon is the same as that we know from our laboratories on Earth.

There are two different approaches to test new physics with black holes and they are normally called, respectively, top-down and bottom-up methods. The top-down strategy is the most natural one: we want to compare the predictions of General Relativity with those from some specific alternative theory in which at least one of the assumptions above does not hold. We can then construct a theoretical model for General Relativity and another theoretical model for the alternative theory. We fit some observational data with the two models and we see if observations prefer one of the two scenarios and can rule out the other one. The bottom-up strategy follows an agnostic and more phenomenological approach. We do not want to test any specific scenario and we construct a model in which possible deviations from the predictions of General Relativity are parametrized. We can then fit the data with this model and estimate the values of these new parameters to verify whether our measurements are consistent with General Relativity or require new physics.

The top-down and bottom-up strategies have their advantages and disadvantages. The main problem with the top-down method is that even if we know the alternative theory we may not know well its predictions. For example, for most gravity theories we do not know their rotating black hole solutions. Often we know their non-rotating solution or some approximate solution valid in the slow-rotating limit, but they are not very useful for our astrophysical tests because astrophysical black holes have normally a non-vanishing spin angular momentum and, actually, we need to study very fast-rotating objects to get convincing results. When the black hole is rotating very fast, the innermost stable circular orbit (ISCO), which under certain circumstances determines the inner edge of the accretion disk, can be close to the black hole event horizon, and this can maximize the impact of the strong gravitational field on the spectrum of the source. In the case of slow-rotating objects, the impact of the relativistic effects on the spectrum is simply too weak for our tests.

In the past years, we have mainly worked on tests of the Kerr metric following the bottom-up approach, so employing some parametric black hole spacetime to measure its deformation parameters. The main reason is that so far we have focused our attention to the development of the models to have the astrophysical part under control, in order to limit the systematic uncertainties, devoting less attention to the exact scenario of new physics. However, once we have a robust model it is relativity straightforward to consider alternative theories, and in our near future plans we expect to follow even the top-down approach to constrain specific theories of gravity.

\subsection{Continuum-fitting method}

The analysis of the thermal spectrum of geometrically thin and optically thick accretion disks around black holes within a relativistic model is normally refereed to as the continuum-fitting method. The technique was first proposed by Shuang-Nan Zhang and collaborators to measure the spin of stellar-mass black holes in X-ray binaries assuming General Relativity~\cite{Zhang:1997dy} and was then developed by the CfA group of Jeff McClintock and Ramesh Narayan~\cite{Li:2004aq,Shafee:2005ef,McClintock:2006xd,McClintock:2011zq,McClintock:2013vwa}. The calculations of the thermal spectrum of a disk depend on the background metric and on the motion of massive particles in the accretion disk and of photons from the emission point in the disk to the detection point, while they are independent of the atomic physics. We can thus test assumptions 1 (Kerr black hole hypothesis) and 2 (Weak Equivalence Principle).

Diego Torres was the first to try to use the continuum-fitting method for testing fundamental physics in~\cite{Torres:2002td}. He calculated thermal spectra of accretion disks around static boson stars and compared the theoretical predictions with that of a Schwarzschild black hole. Thermal spectra of accretion disks around wormholes, brane world black holes, gravastars, etc. were later calculated by other authors~\cite{Pun:2008ae,Harko:2008vy,Pun:2008ua,Harko:2009gc}. The first observational constraints on the Kerr metric with the continuum-fitting method were reported by my group at Fudan University in Ref.~\cite{Kong:2014wha} using the model described in~\cite{Bambi:2011jq,Bambi:2012tg}, where light bending was taken into account.

\begin{figure}[t]
\begin{center}
\includegraphics[width=8.5cm,trim={0cm 0cm 0.5cm 0cm},clip]{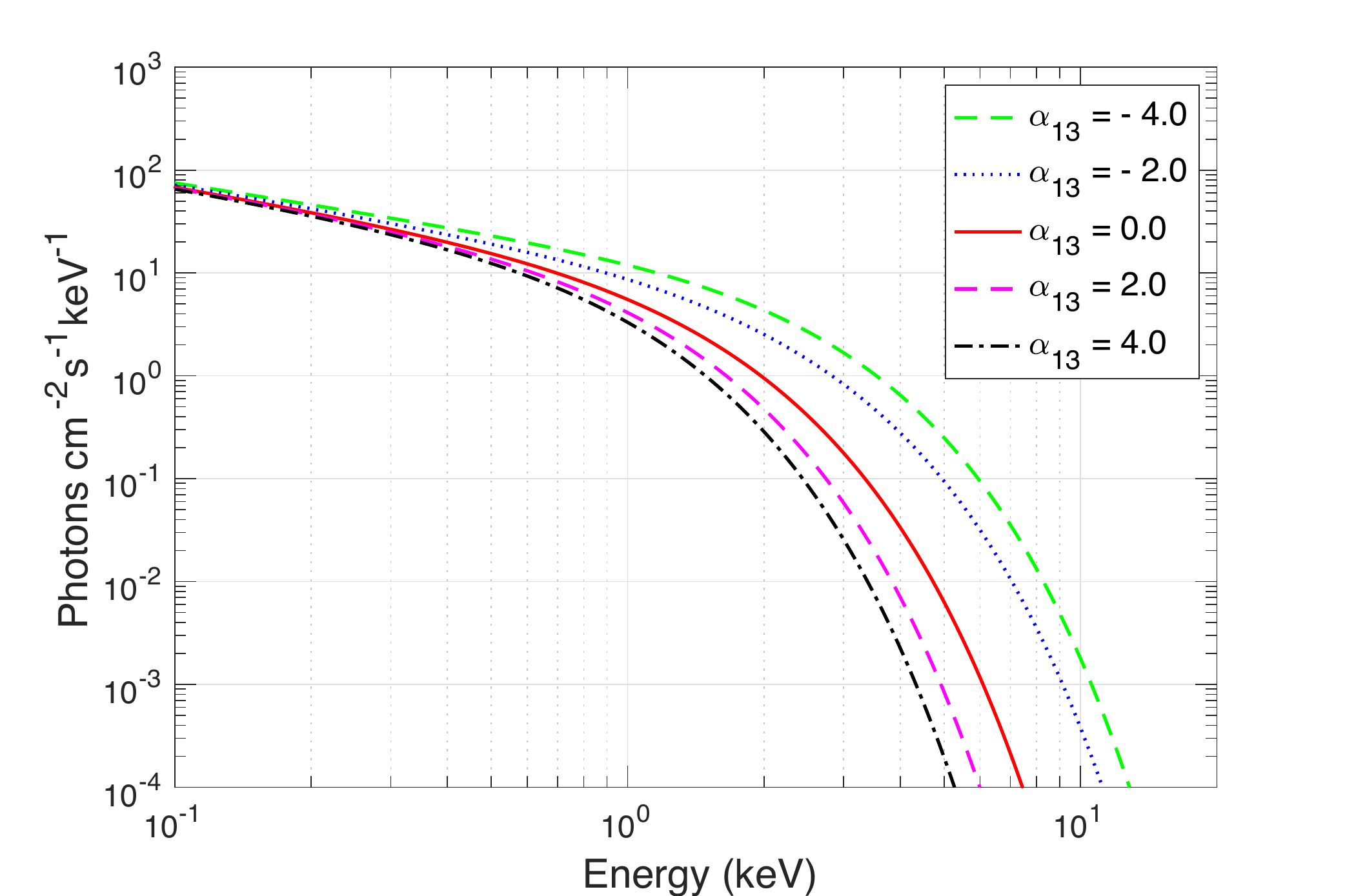}
\end{center}
\vspace{-0.4cm}
\caption{Synthetic thermal spectra of thin disks in the Johannsen spacetime for different values of the deformation parameter $\alpha_{13}$. These spectra are generated with {\tt nkbb} assuming that the black hole mass is $M = 10~M_\odot$, the mass accretion rate is $\dot{M} = 2 \cdot 10^{18}$~g~s$^{-1}$, the black hole distance is $D = 10$~kpc, the inclination angle of the disk is $i = 45^\circ$, and the black hole spin parameter is $a_* = 0.7$. Figure from Ref.~\cite{Zhou:2019fcg}. \label{f-nkbb}}
\end{figure}

More recently, we have developed the XSPEC model {\tt nkbb}~\cite{Zhou:2019fcg}. The first version of the model calculated thermal spectra of accretion disks in the Johannsen metric with the possible non-vanishing deformation parameter $\alpha_{13}$~\cite{Johannsen:2015pca} ($\alpha_{13} = 0$ corresponds to the Kerr metric and we have deviations from the Kerr solution for $\alpha_{13} \neq 0$). Fig.~\ref{f-nkbb} shows the impact of the deformation parameter $\alpha_{13}$ on the thermal spectrum of an accretion disk when the other parameters of the models are fixed. As we can see, the value of $\alpha_{13}$ affects the spectrum of the disk and therefore, modulo degeneracy with other parameters of the model, we can expect to be able to constrain the deformation parameter $\alpha_{13}$ or other deformations from the Kerr spacetime from the analysis of real data. Even if the version described in~\cite{Zhou:2019fcg} employs he Johannsen metric with the deformation parameter $\alpha_{13}$, {\tt nkbb} can be easily modified to test any stationary, axisymmetric, and asymptotical black hole metric with a known analytic expression.

\subsection{X-ray reflection spectroscopy}

X-ray reflection spectroscopy refers to the analysis of the reflection features in the X-ray spectrum of the accretion disk around a black hole. Assuming standard atomic physics, in the rest-frame of the gas in the disk, the reflection spectrum is characterized by narrow fluorescent emission lines below 10~keV (the most prominent feature is often the iron K$\alpha$ complex at 6.4~keV in the case of neutral or weakly ionized iron and up to 6.97~keV in the case of H-like iron ions) and a Compton hump peaked at 20-30~keV~\cite{Ross:2005dm,Garcia:2010iz}. However, relativistic effects in the strong gravitational field of the black hole (Doppler boosting, gravitational redshift, and light-bending) make these lines broadened and skewed in the spectrum detected by a distant observer~\cite{Fabian:1989ej,Laor:1991nc}.

As in the case of the continuum-fitting method, X-ray reflection spectroscopy was proposed and developed in the framework of General Relativity to study the accretion flow around black holes and measure black hole spins~\cite{Brenneman:2006hw}; for a review, see~\cite{Bambi:2012tg} and references therein. Youjun Lu and Diego Torres were the first to study the shape of the iron K$\alpha$ line as a tool to test the spacetime metric around a compact object. In Ref.~\cite{Lu:2002vm}, they calculated the iron line profile emitted from the accretion disk of a static boson star and compared their predictions with the iron line profile expected from the accretion disk of a Schwarzschild black hole. After the work in Ref.~\cite{Lu:2002vm}, other authors calculated the profile of emission lines expected in different black hole spacetimes~\cite{Schee:2008fc,Johannsen:2012ng,Bambi:2012at,Schee:2013bya}.

\begin{figure*}[t]
\begin{center}
\includegraphics[width=8.5cm,trim={0cm 0cm 0cm 0cm},clip]{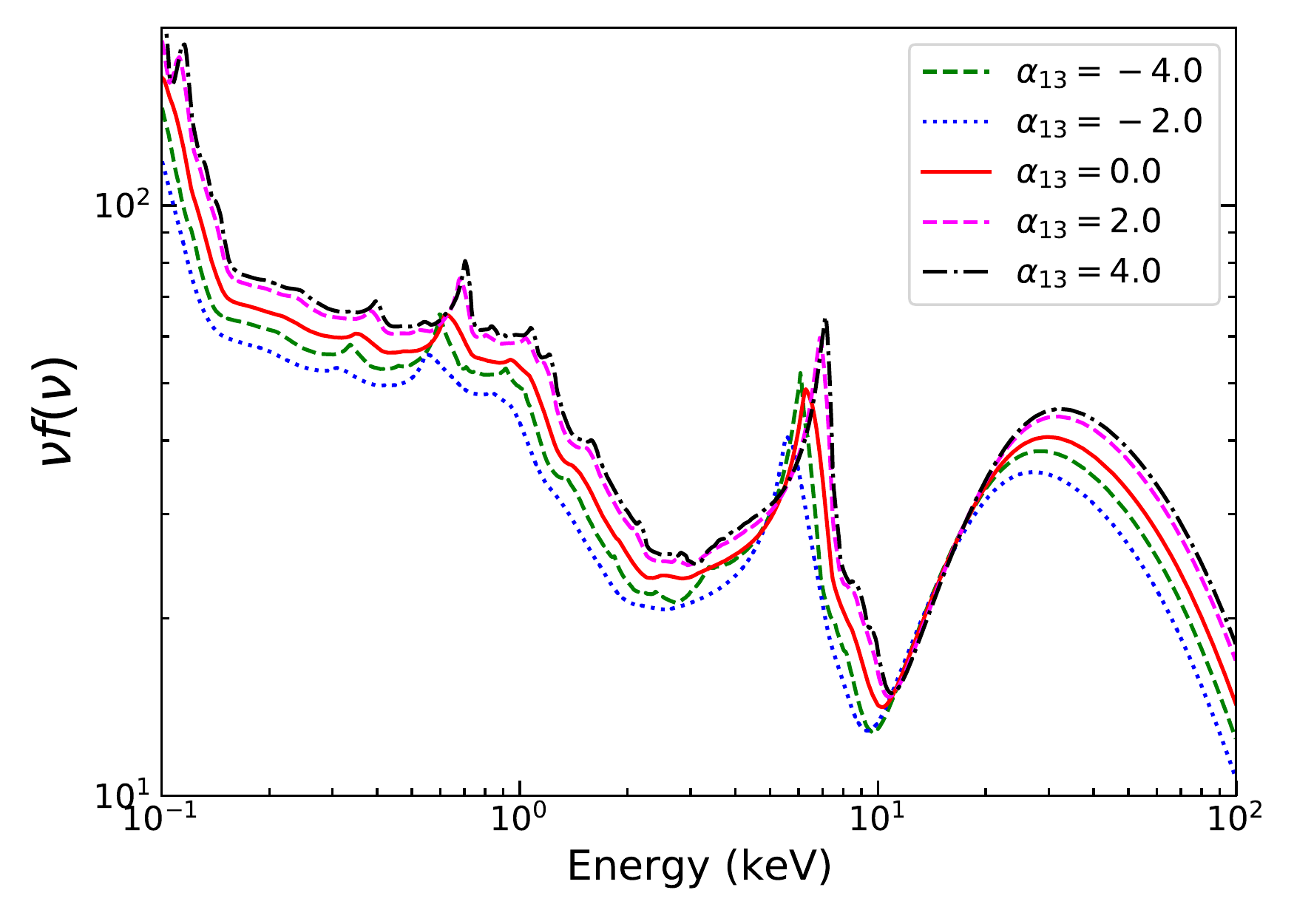}
\hspace{0.3cm}
\includegraphics[width=8.5cm,trim={0cm 0cm 0cm 0cm},clip]{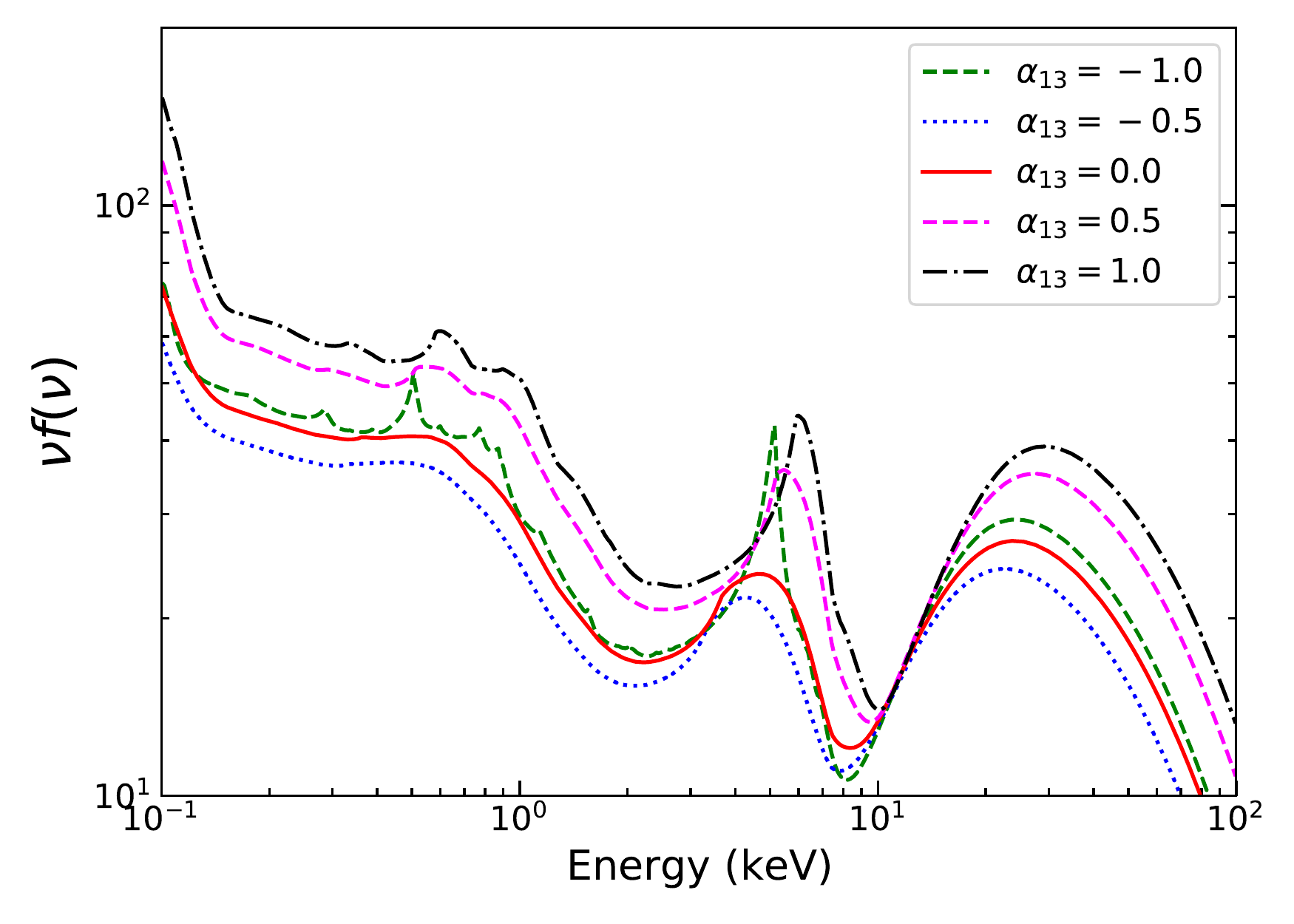}
\end{center}
\vspace{-0.6cm}
\caption{Synthetic reflection spectra of thin disks in the Johannsen spacetime for different values of the deformation parameter $\alpha_{13}$. These spectra are generated with {\tt relxill\_nk} assuming that the incident radiation has a power-law spectrum with photon index $\Gamma = 2$, the emissivity profile is described by a power-law with emissivity index $q = 7$, the ionization parameter of the disk is $\log\xi = 3.1$ ($\xi$ in erg~s~cm$^{-1}$), the disk has Solar iron abundance, the inclination angle of the disk is $i = 45^\circ$, and the black hole spin parameter is $a_* = 0.7$ (left panel) and 0.97 (right panel). \label{f-relxillnk}}
\end{figure*}

In Refs.~\cite{Bambi:2016sac,Abdikamalov:2019yrr}, we presented the first version of the relativistic reflection model {\tt relxill\_nk}, which is an extension of the {\tt relxill} package developed by Thomas Dauser and Javier Garcia~\cite{Dauser:2010ne,Garcia:2013oma,Garcia:2013lxa}. As {\tt nkbb}, the default version of {\tt relxill\_nk} employs the Johannsen metric with the possible non-vanishing deformation parameter $\alpha_{13}$~\cite{Johannsen:2015pca}, but modified versions of {\tt relxill\_nk} have been used to test black hole solutions in conformal gravity~\cite{Zhou:2018bxk,Zhou:2019hqk}, in Kaluza-Klein models~\cite{Zhu:2020cfn}, in asymptotically safe quantum gravity~\cite{Zhou:2020eth}, and in Einstein-Maxwell dilaton-axion gravity~\cite{Tripathi:2021rwb}, as in all these theories we know the analytic expression of the metric of rotating black holes. {\tt relxill\_nk} has also been used for testing the Weak Equivalence Principle~\cite{Roy:2021pns}, while we have not used our model for testing the atomic physics in the strong gravitational fields of black holes.

The {\tt relxill\_nk} package is under development. It has already the possibility of calculating the reflection spectrum for a disk of finite thickness~\cite{Abdikamalov:2020oci} and including a simple ionization parameter radial profile over the disk~\cite{Abdikamalov:2021rty}. The version with a non-trivial electron density profile and a self-consistent ionization profile should be released soon, while we are already working to include in the calculations the effect of the returning radiation, namely the radiation emitted by the disk and returning to the disk because of the strong light bending.

Fig.~\ref{f-relxillnk} shows synthetic reflection spectra for different values of the Johannsen deformation parameter $\alpha_{13}$. As in the case of Fig.~\ref{f-nkbb}, we clearly see that the value of $\alpha_{13}$ has an impact on the shape of the spectrum of the disk and therefore, modulo degeneracy with other parameters of the model, we can expect that the analysis of the reflection features in the X-ray spectrum of a black hole can lead to constrain the value of the deformation parameter $\alpha_{13}$.


\section{Results}

In the past few years, we have used {\tt nkbb} and {\tt relxill\_nk} to test stellar-mass black holes in X-ray binary systems and supermassive black holes in AGN using data from \textsl{NuSTAR}, \textsl{RXTE}, \textsl{Suzaku}, \textsl{Swift}, and \textsl{XMM-Newton}. Our constraints on the Johannsen deformation parameter $\alpha_{13}$ are summarized in Tab.~\ref{t-bhb} (stellar-mass black holes) and Tab.~\ref{t-agn} (supermassive black holes). We note that some of the constraints on $\alpha_{13}$ in Tab.~\ref{t-bhb} and Tab.~\ref{t-agn} do not perfectly match with the measurements on $\alpha_{13}$ reported in the original papers (even if they are always consistent with them). This is because we have re-analyzed some observations with a more recent version of our models, and the constraints reported in Tab.~\ref{t-bhb} and Tab.~\ref{t-agn} should be regarded as more accurate than those in the original papers. Tab.~\ref{t-bhb} also shows the constraints on $\alpha_{13}$ inferred from gravitational wave data in GWTC-1 and Tab.~\ref{t-agn} reports the constraint on $\alpha_{13}$ from the observation of the shadow of the black hole in the galaxy M87.

All current measurements of $\alpha_{13}$ are consistent with $\alpha_{13} = 0$, as it is requested by General Relativity. However, the precision and the accuracy of different measurements can be quite different. For stellar-mass black holes, the most robust and precise constraints are those obtained from EXO~1846--031, GRS~1716--249, GRS~1915+105, and GX~339--4. These are all very bright sources and, for three of them, the constraint is obtained by a combined analysis of the thermal spectrum and the reflection features. In the case of supermassive black hole, the best constraint on $\alpha_{13}$ is obtained from MCG--6--30--15, where we have simultaneous observations of \textsl{NuSTAR} and \textsl{XMM-Newton}. The measurements of $\alpha_{13}$ in Tab.~\ref{t-agn} with \textsl{Suzaku} data have to be taken with more caution as they were obtained only analyzing the 1-10~keV band of the spectrum, which includes the iron line and does not include the Compton hump.

\begin{table*}
\centering
{\renewcommand{\arraystretch}{1.3}
\begin{tabular}{lcccc}
\hline\hline
Source &  \hspace{1.0cm} Data \hspace{1.0cm}  & \hspace{0.5cm} $\alpha_{13}$ (3-$\sigma$) \hspace{0.5cm} & \hspace{0.5cm} Method \hspace{0.5cm} & \hspace{0.5cm} Main Reference \hspace{0.5cm} \\
\hline\hline
4U~1630--472 & \textsl{NuSTAR} & $-0.03_{-0.18}^{+0.63}$ & reflection & \cite{Tripathi:2020yts} \\
Cygnus~X-1 & \textsl{Suzaku} & $-0.2_{-0.8}^{+0.5}$ & reflection & \cite{Zhang:2020qbx} \\ 
EXO~1846--031 & \textsl{NuSTAR} & $-0.03_{-0.18}^{+0.17}$ & reflection & \cite{Tripathi:2020yts} \\
GRS~1716--249 & \textsl{NuSTAR}+\textsl{Swift} & $0.09_{-0.26}^{+0.02}$ & CFM + reflection & \cite{zuobin} \\
GRS~1739--278 & \textsl{NuSTAR} & $-0.3_{-0.5}^{+0.6}$ & reflection & \cite{Tripathi:2020yts} \\
GRS~1915+105 & \textsl{Suzaku} & $0.00_{-0.26}^{+0.17}$ & reflection & \cite{Zhang:2019ldz} \\
& \textsl{RXTE}+\textsl{Suzaku} & $0.12_{-0.27}^{+0.02}$ & CFM + reflection & \cite{ashutosh} \\
GS~1354--645 & \textsl{NuSTAR} & $0.0_{-0.9}^{+0.6}$ & reflection & \cite{Xu:2018lom} \\
GW150914 & GWTC-1 & $-0.9 \pm 1.3$ & GW & \cite{Cardenas-Avendano:2019zxd} \\
GW151226 & GWTC-1 & $0.0 \pm 1.2$ & GW & \cite{Cardenas-Avendano:2019zxd} \\
GW170104 & GWTC-1 & $1.7 \pm 3.1$ & GW & \cite{Cardenas-Avendano:2019zxd} \\
GW170608 & GWTC-1 & $-0.1 \pm 0.8$ & GW & \cite{Cardenas-Avendano:2019zxd} \\
GW170814 & GWTC-1 & $-0.2 \pm 1.4$ & GW & \cite{Cardenas-Avendano:2019zxd} \\
GX~339--4 & \textsl{NuSTAR}+\textsl{Swift} & $-0.02_{-0.14}^{+0.03}$ & CFM + reflection & \cite{Tripathi:2020dni} \\
LMC~X-1 & \textsl{RXTE} & $< 0.4$ & CFM & \cite{Tripathi:2020qco} \\
Swift~J1658--4242 & \textsl{NuSTAR}+\textsl{Swift} & $0.0_{-1.0}^{+1.2}$ & reflection & \cite{Tripathi:2020yts} \\
\hline\hline
\end{tabular}}
\caption{Summary of the 3-$\sigma$ constraints on the Johannsen deformation parameter $\alpha_{13}$ from stellar-mass black holes. CFM = continuum-fitting method ({\tt nkbb}); reflection = X-ray reflection spectroscopy ({\tt relxill\_nk}); GW = gravitational waves (inspiral phase). \label{t-bhb}}
\end{table*}

\begin{table*}
\centering
{\renewcommand{\arraystretch}{1.3}
\begin{tabular}{lcccc}
\hline\hline
Source &  \hspace{1.0cm} Data \hspace{1.0cm}  & \hspace{0.5cm} $\alpha_{13}$ (3-$\sigma$) \hspace{0.5cm} & \hspace{0.5cm} Method \hspace{0.5cm} & \hspace{0.5cm} Main Reference \hspace{0.5cm} \\
\hline\hline
1H0419--577 & \textsl{Suzaku} & $0.00_{-0.35}^{+0.12}$ & reflection & \cite{Tripathi:2019bya} \\
1H0707--495 & \textsl{NuSTAR}+\textsl{Swift} & $-2.0 < \alpha_{13} < 0.6$ & reflection & \cite{Cao:2017kdq} \\ 
Ark~120 & \textsl{Suzaku} & $0.00_{-0.37}^{+0.08}$ & reflection & \cite{Tripathi:2019bya} \\
Ark~564 & \textsl{Suzaku} & $-0.2_{-0.8}^{+0.4}$ & reflection & \cite{Tripathi:2018bbu} \\
Fairall~9 & \textsl{NuSTAR}+\textsl{XMM} & $-1.4 < \alpha_{13} < 0.4$ & reflection & \cite{Liu:2020fpv} \\
M87$^\star$ & \textsl{EHT} & $-3.6 < \alpha_{13} < 5.9 {}^\dag$ & VLBI & \cite{Psaltis:2020lvx} \\
MCG--6--30--15 & \textsl{NuSTAR}+\textsl{XMM} & $0.00_{-0.44}^{+0.15}$ & reflection & \cite{Tripathi:2018lhx} \\
Mrk~335 & \textsl{Suzaku} & $-3.0 < \alpha_{13} < 0.5$ & reflection & \cite{Choudhury:2018zmf} \\
PKS~0558--504 & \textsl{Suzaku} & $-0.7_{-1.5}^{+1.4}$ & reflection & \cite{Tripathi:2019bya} \\
Swift~J0501.9--3239 & \textsl{Suzaku} & $0.00_{-0.66}^{+0.11}$ & reflection & \cite{Tripathi:2019bya} \\
Ton~S180 & \textsl{Suzaku} & $0.01_{-0.50}^{+0.07}$ & reflection & \cite{Tripathi:2019bya} \\
\hline\hline
\end{tabular}}
\caption{Summary of the 3-$\sigma$ constraints on the Johannsen deformation parameter $\alpha_{13}$ from supermassive black holes. reflection = X-ray reflection spectroscopy ({\tt relxill\_nk}); VLBI = very long baseline interferometry. $^\dag$ 1-$\sigma$ constraint. \label{t-agn}}
\end{table*}


\section{Concluding remarks}

In the past years, we have developed two XSPEC models to test General Relativity with black hole X-ray data: the multi-temperature blackbody model {\tt nkbb} and the relativistic reflection model {\tt relxill\_nk}. They currently represent the state-of-the-art in this research field. We have already used these models to analyze a number of X-ray data of stellar-mass black holes in X-ray binary systems and of supermassive black holes in AGN, mainly to test the Kerr metric around these compact objects.

Our near future plans (next 2-3~years) can be summarized as follows.
\begin{enumerate}
\item Development of the models. We are going to further improve {\tt nkbb} and {\tt relxill\_nk}, which we expect to be strictly necessary in view of very high-quality data that will be available with the next generation of X-ray missions; e.g., \textsl{Athena}~\cite{Nandra:2013jka} and \textsl{eXTP}~\cite{Zhang:2016ach}. Our current priority is to include the returning radiation in the spectra calculated by {\tt relxill\_nk}. This is the radiation emitted by the disk and returning to the disk because of the strong light bending near the black hole; see, e.g., \cite{Riaz:2020zqb} and references therein. 
\item As of now, we have mainly focused our efforts on the development of the astrophysical part of the codes, devoting less attention to testing specific gravity models. For the future, we plan to extend our studies to test specific theories of gravity. This will require to develop {\tt nkbb} and {\tt relxill\_nk} to work with numerical metrics, as rotating black hole solutions beyond General Relativity are rarely known in analytic form and often only numerical metrics are available.
\item Up to now we have mainly considered tests of the Kerr metric. Once again, the main reason is that we have devoted most of our efforts to develop the astrophysical parts of {\tt nkbb} and {\tt relxill\_nk} in order to have the systematics under control, while we have somewhat neglected the rich phenomenology of theories beyond General Relativity. Our plan is thus to use black hole X-ray data even to test the geodesic motions of massive and massless particles and atomic physics in the strong gravitational field near the black hole event horizon.
\end{enumerate}


\vspace{0.5cm}

{\bf Acknowledgments --}
This work was supported by the Innovation Program of the Shanghai Municipal Education Commission, Grant No.~2019-01-07-00-07-E00035, the National Natural Science Foundation of China (NSFC), Grant No.~11973019, and Fudan University, Grant No.~JIH1512604.



\begin{thebibliography}{99}

\bibitem{Einstein:1916vd}
A.~Einstein,
{\it The Foundation of the General Theory of Relativity},
Annalen Phys. \textbf{49}, 769-822 (1916).

\bibitem{Will:2014kxa}
C.~M.~Will,
{\it The Confrontation between General Relativity and Experiment},
Living Rev. Rel. \textbf{17}, 4 (2014)
[arXiv:1403.7377 [gr-qc]].

\bibitem{Jain:2010ka}
B.~Jain and J.~Khoury,
{\it Cosmological Tests of Gravity},
Annals Phys. \textbf{325}, 1479-1516 (2010)
[arXiv:1004.3294 [astro-ph.CO]].

\bibitem{Koyama:2015vza}
K.~Koyama,
{\it Cosmological Tests of Modified Gravity},
Rept. Prog. Phys. \textbf{79}, 046902 (2016)
[arXiv:1504.04623 [astro-ph.CO]].

\bibitem{Ferreira:2019xrr}
P.~G.~Ferreira,
{\it Cosmological Tests of Gravity},
Ann. Rev. Astron. Astrophys. \textbf{57}, 335-374 (2019)
[arXiv:1902.10503 [astro-ph.CO]].

\bibitem{Bambi:2015kza}
C.~Bambi,
{\it Testing black hole candidates with electromagnetic radiation},
Rev. Mod. Phys. \textbf{89}, 025001 (2017)
[arXiv:1509.03884 [gr-qc]].

\bibitem{Yagi:2016jml}
K.~Yagi and L.~C.~Stein,
{\it Black Hole Based Tests of General Relativity},
Class. Quant. Grav. \textbf{33}, 054001 (2016)
[arXiv:1602.02413 [gr-qc]].

\bibitem{TheLIGOScientific:2016src}
B.~P.~Abbott \textit{et al.} [LIGO Scientific and Virgo],
{\it Tests of general relativity with GW150914},
Phys. Rev. Lett. \textbf{116}, 221101 (2016)
[erratum: Phys. Rev. Lett. \textbf{121}, 129902 (2018)]
[arXiv:1602.03841 [gr-qc]].

\bibitem{Yunes:2016jcc}
N.~Yunes, K.~Yagi and F.~Pretorius,
{\it Theoretical Physics Implications of the Binary Black-Hole Mergers GW150914 and GW151226},
Phys. Rev. D \textbf{94}, 084002 (2016)
[arXiv:1603.08955 [gr-qc]].

\bibitem{LIGOScientific:2019fpa}
B.~P.~Abbott \textit{et al.} [LIGO Scientific and Virgo],
{\it Tests of General Relativity with the Binary Black Hole Signals from the LIGO-Virgo Catalog GWTC-1},
Phys. Rev. D \textbf{100}, 104036 (2019)
[arXiv:1903.04467 [gr-qc]].

\bibitem{Davoudiasl:2019nlo}
H.~Davoudiasl and P.~B.~Denton,
{\it Ultralight Boson Dark Matter and Event Horizon Telescope Observations of M87*},
Phys. Rev. Lett. \textbf{123}, 021102 (2019)
[arXiv:1904.09242 [astro-ph.CO]].

\bibitem{Bambi:2019tjh}
C.~Bambi, K.~Freese, S.~Vagnozzi and L.~Visinelli,
{\it Testing the rotational nature of the supermassive object M87* from the circularity and size of its first image},
Phys. Rev. D \textbf{100}, 044057 (2019)
[arXiv:1904.12983 [gr-qc]].

\bibitem{Psaltis:2020lvx}
D.~Psaltis \textit{et al.} [Event Horizon Telescope],
{\it Gravitational Test Beyond the First Post-Newtonian Order with the Shadow of the M87 Black Hole},
Phys. Rev. Lett. \textbf{125}, 141104 (2020)
[arXiv:2010.01055 [gr-qc]].

\bibitem{Cao:2017kdq}
Z.~Cao, S.~Nampalliwar, C.~Bambi, T.~Dauser and J.~A.~Garcia,
{\it Testing general relativity with the reflection spectrum of the supermassive black hole in 1H0707$-$495},
Phys. Rev. Lett. \textbf{120}, 051101 (2018)
[arXiv:1709.00219 [gr-qc]].

\bibitem{Tripathi:2018lhx}
A.~Tripathi, S.~Nampalliwar, A.~B.~Abdikamalov, D.~Ayzenberg, C.~Bambi, T.~Dauser, J.~A.~Garcia and A.~Marinucci,
{\it Toward Precision Tests of General Relativity with Black Hole X-Ray Reflection Spectroscopy},
Astrophys. J. \textbf{875}, 56 (2019)
[arXiv:1811.08148 [gr-qc]].

\bibitem{Tripathi:2020dni}
A.~Tripathi, A.~B.~Abdikamalov, D.~Ayzenberg, C.~Bambi, V.~Grinberg and M.~Zhou,
{\it Testing the Kerr Black Hole Hypothesis with GX 339\textendash{}4 by a Combined Analysis of Its Thermal Spectrum and Reflection Features},
Astrophys. J. \textbf{907}, 31 (2021)
[arXiv:2010.13474 [astro-ph.HE]].

\bibitem{Carter:1971zc}
B.~Carter,
{\it Axisymmetric Black Hole Has Only Two Degrees of Freedom},
Phys. Rev. Lett. \textbf{26}, 331-333 (1971).

\bibitem{Robinson:1975bv}
D.~C.~Robinson,
{\it Uniqueness of the Kerr black hole},
Phys. Rev. Lett. \textbf{34}, 905-906 (1975).

\bibitem{Chrusciel:2012jk}
P.~T.~Chrusciel, J.~Lopes Costa and M.~Heusler,
{\it Stationary Black Holes: Uniqueness and Beyond},
Living Rev. Rel. \textbf{15}, 7 (2012)
[arXiv:1205.6112 [gr-qc]].

\bibitem{Kerr:1963ud}
R.~P.~Kerr,
{\it Gravitational field of a spinning mass as an example of algebraically special metrics},
Phys. Rev. Lett. \textbf{11}, 237-238 (1963).

\bibitem{Bambi:2008hp}
C.~Bambi, A.~D.~Dolgov and A.~A.~Petrov,
{\it Black holes as antimatter factories},
JCAP \textbf{09}, 013 (2009)
[arXiv:0806.3440 [astro-ph]].

\bibitem{Bambi:2014koa}
C.~Bambi, D.~Malafarina and N.~Tsukamoto,
{\it Note on the effect of a massive accretion disk in the measurements of black hole spins},
Phys. Rev. D \textbf{89}, 127302 (2014)
[arXiv:1406.2181 [gr-qc]].

\bibitem{Bambi:2017khi}
C.~Bambi,
{\it Black Holes: A Laboratory for Testing Strong Gravity} (Springer Singapore, 2017),
doi:10.1007/978-981-10-4524-0

\bibitem{Dvali:2012rt}
G.~Dvali and C.~Gomez,
{\it Black Hole's 1/N Hair},
Phys. Lett. B \textbf{719}, 419-423 (2013)
[arXiv:1203.6575 [hep-th]].

\bibitem{Giddings:2014ova}
S.~B.~Giddings,
{\it Possible observational windows for quantum effects from black holes},
Phys. Rev. D \textbf{90}, 124033 (2014)
[arXiv:1406.7001 [hep-th]].

\bibitem{Giddings:2017jts}
S.~B.~Giddings,
{\it Astronomical tests for quantum black hole structure},
Nature Astron. \textbf{1}, 0067 (2017)
[arXiv:1703.03387 [gr-qc]].

\bibitem{Herdeiro:2014goa}
C.~A.~R.~Herdeiro and E.~Radu,
{\it Kerr black holes with scalar hair},
Phys. Rev. Lett. \textbf{112}, 221101 (2014)
[arXiv:1403.2757 [gr-qc]].

\bibitem{Kleihaus:2011tg}
B.~Kleihaus, J.~Kunz and E.~Radu,
{\it Rotating Black Holes in Dilatonic Einstein-Gauss-Bonnet Theory},
Phys. Rev. Lett. \textbf{106}, 151104 (2011)
[arXiv:1101.2868 [gr-qc]].

\bibitem{Ayzenberg:2014aka}
D.~Ayzenberg and N.~Yunes,
{\it Slowly-Rotating Black Holes in Einstein-Dilaton-Gauss-Bonnet Gravity: Quadratic Order in Spin Solutions},
Phys. Rev. D \textbf{90}, 044066 (2014)
[erratum: Phys. Rev. D \textbf{91}, 069905 (2015)]
[arXiv:1405.2133 [gr-qc]].

\bibitem{Bambi:2020jpe}
C.~Bambi \textit{et al.},
{\it Towards precision measurements of accreting black holes using X-ray reflection spectroscopy},
[arXiv:2011.04792 [astro-ph.HE]].  

\bibitem{Zhang:1997dy}
S.~N.~Zhang, W.~Cui and W.~Chen,
{\it Black hole spin in X-ray binaries: Observational consequences},
Astrophys. J. Lett. \textbf{482}, L155 (1997)
[arXiv:astro-ph/9704072 [astro-ph]].

\bibitem{Li:2004aq}
L.~X.~Li, E.~R.~Zimmerman, R.~Narayan and J.~E.~McClintock,
{\it Multi-temperature blackbody spectrum of a thin accretion disk around a Kerr black hole: Model computations and comparison with observations},
Astrophys. J. Suppl. \textbf{157}, 335-370 (2005)
[arXiv:astro-ph/0411583 [astro-ph]].

\bibitem{Shafee:2005ef}
R.~Shafee, J.~E.~McClintock, R.~Narayan, S.~W.~Davis, L.~X.~Li and R.~A.~Remillard,
{\it Estimating the spin of stellar-mass black holes via spectral fitting of the x-ray continuum},
Astrophys. J. Lett. \textbf{636}, L113-L116 (2006)
[arXiv:astro-ph/0508302 [astro-ph]].

\bibitem{McClintock:2006xd}
J.~E.~McClintock, R.~Shafee, R.~Narayan, R.~A.~Remillard, S.~W.~Davis and L.~X.~Li,
{\it The Spin of the Near-Extreme Kerr Black Hole GRS 1915+105},
Astrophys. J. \textbf{652}, 518-539 (2006)
[arXiv:astro-ph/0606076 [astro-ph]].

\bibitem{McClintock:2011zq}
J.~E.~McClintock, R.~Narayan, S.~W.~Davis, L.~Gou, A.~Kulkarni, J.~A.~Orosz, R.~F.~Penna, R.~A.~Remillard and J.~F.~Steiner,
{\it Measuring the Spins of Accreting Black Holes},
Class. Quant. Grav. \textbf{28}, 114009 (2011)
[arXiv:1101.0811 [astro-ph.HE]].

\bibitem{McClintock:2013vwa}
J.~E.~McClintock, R.~Narayan and J.~F.~Steiner,
{\it Black Hole Spin via Continuum Fitting and the Role of Spin in Powering Transient Jets},
Space Sci. Rev. \textbf{183}, 295-322 (2014)
[arXiv:1303.1583 [astro-ph.HE]].

\bibitem{Torres:2002td}
D.~F.~Torres,
{\it Accretion disc onto a static nonbaryonic compact object},
Nucl. Phys. B \textbf{626}, 377-394 (2002)
[arXiv:hep-ph/0201154 [hep-ph]].

\bibitem{Pun:2008ae}
C.~S.~J.~Pun, Z.~Kovacs and T.~Harko,
{\it Thin accretion disks in f(R) modified gravity models},
Phys. Rev. D \textbf{78}, 024043 (2008)
[arXiv:0806.0679 [gr-qc]].

\bibitem{Harko:2008vy}
T.~Harko, Z.~Kovacs and F.~S.~N.~Lobo,
{\it Electromagnetic signatures of thin accretion disks in wormhole geometries},
Phys. Rev. D \textbf{78}, 084005 (2008)
[arXiv:0808.3306 [gr-qc]].

\bibitem{Pun:2008ua}
C.~S.~J.~Pun, Z.~Kovacs and T.~Harko,
{\it Thin accretion disks onto brane world black holes},
Phys. Rev. D \textbf{78}, 084015 (2008)
[arXiv:0809.1284 [gr-qc]].

\bibitem{Harko:2009gc}
T.~Harko, Z.~Kovacs and F.~S.~N.~Lobo,
{\it Can accretion disk properties distinguish gravastars from black holes?},
Class. Quant. Grav. \textbf{26}, 215006 (2009)
[arXiv:0905.1355 [gr-qc]].

\bibitem{Kong:2014wha}
L.~Kong, Z.~Li and C.~Bambi,
{\it Constraints on the spacetime geometry around 10 stellar-mass black hole candidates from the disk's thermal spectrum},
Astrophys. J. \textbf{797}, 78 (2014)
[arXiv:1405.1508 [gr-qc]].

\bibitem{Bambi:2011jq}
C.~Bambi and E.~Barausse,
{\it Constraining the quadrupole moment of stellar-mass black-hole candidates with the continuum fitting method},
Astrophys. J. \textbf{731}, 121 (2011)
[arXiv:1012.2007 [gr-qc]].

\bibitem{Bambi:2012tg}
C.~Bambi,
{\it A code to compute the emission of thin accretion disks in non-Kerr space-times and test the nature of black hole candidates},
Astrophys. J. \textbf{761}, 174 (2012)
[arXiv:1210.5679 [gr-qc]].

\bibitem{Zhou:2019fcg}
M.~Zhou, A.~B.~Abdikamalov, D.~Ayzenberg, C.~Bambi, H.~Liu and S.~Nampalliwar,
{\it XSPEC model for testing the Kerr black hole hypothesis using the continuum-fitting method},
Phys. Rev. D \textbf{99}, 104031 (2019)
[arXiv:1903.09782 [gr-qc]].

\bibitem{Johannsen:2015pca}
T.~Johannsen,
{\it Regular Black Hole Metric with Three Constants of Motion},
Phys. Rev. D \textbf{88}, 044002 (2013)
[arXiv:1501.02809 [gr-qc]].

\bibitem{Ross:2005dm}
R.~R.~Ross and A.~C.~Fabian,
{\it A Comprehensive range of x-ray ionized reflection models},
Mon. Not. Roy. Astron. Soc. \textbf{358}, 211-216 (2005)
[arXiv:astro-ph/0501116 [astro-ph]].  
  
\bibitem{Garcia:2010iz}
J.~Garcia and T.~Kallman,
{\it X-ray reflected spectra from accretion disk models. I. Constant density atmospheres},
Astrophys. J. \textbf{718}, 695 (2010)
[arXiv:1006.0485 [astro-ph.HE]].  

\bibitem{Fabian:1989ej}
A.~C.~Fabian, M.~J.~Rees, L.~Stella and N.~E.~White,
{\it X-ray fluorescence from the inner disc in Cygnus X-1},
Mon. Not. Roy. Astron. Soc. \textbf{238}, 729-736 (1989).
  
\bibitem{Laor:1991nc}
A.~Laor,
{\it Line profiles from a disk around a rotating black hole},
Astrophys. J. \textbf{376}, 90 (1991). 

\bibitem{Brenneman:2006hw}
L.~W.~Brenneman and C.~S.~Reynolds,
{\it Constraining Black Hole Spin Via X-ray Spectroscopy},
Astrophys. J. \textbf{652}, 1028-1043 (2006)
[arXiv:astro-ph/0608502 [astro-ph]].

\bibitem{Lu:2002vm}
Y.~Lu and D.~F.~Torres,
{\it The relativistic iron k-alpha line from an accretion disc onto a static non-baryonic compact object},
Int. J. Mod. Phys. D \textbf{12}, 63-78 (2003)
[arXiv:astro-ph/0205418 [astro-ph]].

\bibitem{Schee:2008fc}
J.~Schee and Z.~Stuchlik,
{\it Profiles of emission lines generated by rings orbiting braneworld Kerr black holes},
Gen. Rel. Grav. \textbf{41}, 1795-1818 (2009)
[arXiv:0812.3017 [astro-ph]].

\bibitem{Johannsen:2012ng}
T.~Johannsen and D.~Psaltis,
{\it Testing the No-Hair Theorem with Observations in the Electromagnetic Spectrum. IV. Relativistically Broadened Iron Lines},
Astrophys. J. \textbf{773}, 57 (2013)
[arXiv:1202.6069 [astro-ph.HE]].

\bibitem{Bambi:2012at}
C.~Bambi,
{\it Testing the space-time geometry around black hole candidates with the analysis of the broad K$\alpha$ iron line},
Phys. Rev. D \textbf{87}, 023007 (2013)
[arXiv:1211.2513 [gr-qc]].

\bibitem{Schee:2013bya}
J.~Schee and Z.~Stuchlik,
{\it Profiled spectral lines generated in the field of Kerr superspinars},
JCAP \textbf{04}, 005 (2013).

\bibitem{Bambi:2016sac}
C.~Bambi, A.~Cardenas-Avendano, T.~Dauser, J.~A.~Garcia and S.~Nampalliwar,
{\it Testing the Kerr black hole hypothesis using X-ray reflection spectroscopy},
Astrophys. J. \textbf{842}, 76 (2017)
[arXiv:1607.00596 [gr-qc]].

\bibitem{Abdikamalov:2019yrr}
A.~B.~Abdikamalov, D.~Ayzenberg, C.~Bambi, T.~Dauser, J.~A.~Garcia and S.~Nampalliwar,
{\it Public Release of RELXILL\_NK: A Relativistic Reflection Model for Testing Einstein\textquoteright{}s Gravity},
Astrophys. J. \textbf{878}, 91 (2019)
[arXiv:1902.09665 [gr-qc]].

\bibitem{Dauser:2010ne}
T.~Dauser, J.~Wilms, C.~S.~Reynolds and L.~W.~Brenneman,
{\it Broad emission lines for negatively spinning black holes},
Mon. Not. Roy. Astron. Soc. \textbf{409}, 1534 (2010)
[arXiv:1007.4937 [astro-ph.HE]].

\bibitem{Garcia:2013oma}
J.~Garcia, T.~Dauser, C.~S.~Reynolds, T.~R.~Kallman, J.~E.~McClintock, J.~Wilms and W.~Eikmann,
{\it X-ray reflected spectra from accretion disk models. III. A complete grid of ionized reflection calculations},
Astrophys. J. \textbf{768}, 146 (2013)
[arXiv:1303.2112 [astro-ph.HE]].

\bibitem{Garcia:2013lxa}
J.~Garc\'\i{}a, T.~Dauser, A.~Lohfink, T.~R.~Kallman, J.~Steiner, J.~E.~McClintock, L.~Brenneman, J.~Wilms, W.~Eikmann and C.~S.~Reynolds, \textit{et al.}
{\it Improved Reflection Models of Black-Hole Accretion Disks: Treating the Angular Distribution of X-rays},
Astrophys. J. \textbf{782}, 76 (2014)
[arXiv:1312.3231 [astro-ph.HE]].

\bibitem{Zhou:2018bxk}
M.~Zhou, Z.~Cao, A.~Abdikamalov, D.~Ayzenberg, C.~Bambi, L.~Modesto and S.~Nampalliwar,
{\it Testing conformal gravity with the supermassive black hole in 1H0707-495},
Phys. Rev. D \textbf{98}, 024007 (2018)
[arXiv:1803.07849 [gr-qc]].

\bibitem{Zhou:2019hqk}
M.~Zhou, A.~B.~Abdikamalov, D.~Ayzenberg, C.~Bambi, L.~Modesto, S.~Nampalliwar and Y.~Xu,
{\it Singularity-free black holes in conformal gravity: New observational constraints},
EPL \textbf{125}, 30002 (2019)
[arXiv:2003.03738 [gr-qc]].

\bibitem{Zhu:2020cfn}
J.~Zhu, A.~B.~Abdikamalov, D.~Ayzenberg, M.~Azreg-Ainou, C.~Bambi, M.~Jamil, S.~Nampalliwar, A.~Tripathi and M.~Zhou,
{\it X-ray reflection spectroscopy with Kaluza-Klein black holes},
Eur. Phys. J. C \textbf{80}, 622 (2020)
[arXiv:2005.00184 [gr-qc]].

\bibitem{Zhou:2020eth}
B.~Zhou, A.~B.~Abdikamalov, D.~Ayzenberg, C.~Bambi, S.~Nampalliwar and A.~Tripathi,
{\it Shining X-rays on asymptotically safe quantum gravity},
JCAP \textbf{01}, 047 (2021)
[arXiv:2005.12958 [astro-ph.HE]].

\bibitem{Tripathi:2021rwb}
A.~Tripathi, B.~Zhou, A.~B.~Abdikamalov, D.~Ayzenberg and C.~Bambi,
{\it Constraints on Einstein-Maxwell dilaton-axion gravity from X-ray reflection spectroscopy},
[arXiv:2103.07593 [astro-ph.HE]].

\bibitem{Roy:2021pns}
R.~Roy, A.~B.~Abdikamalov, D.~Ayzenberg, C.~Bambi, S.~Riaz and A.~Tripathi,
{\it Testing the Weak Equivalence Principle near black holes},
[arXiv:2103.08978 [astro-ph.HE]].

\bibitem{Abdikamalov:2020oci}
A.~B.~Abdikamalov, D.~Ayzenberg, C.~Bambi, T.~Dauser, J.~A.~Garcia, S.~Nampalliwar, A.~Tripathi and M.~Zhou,
{\it Testing the Kerr black hole hypothesis using X-ray reflection spectroscopy and a thin disk model with finite thickness},
Astrophys. J. \textbf{899}, 80 (2020)
[arXiv:2003.09663 [astro-ph.HE]].

\bibitem{Abdikamalov:2021rty}
A.~B.~Abdikamalov, D.~Ayzenberg, C.~Bambi, H.~Liu and Y.~Zhang,
{\it Implementation of a radial disk ionization profile in the relxill\_nk model},
Phys. Rev. D \textbf{103}, 103023 (2021)
[arXiv:2101.10100 [astro-ph.HE]].

\bibitem{Tripathi:2020yts}
A.~Tripathi, Y.~Zhang, A.~B.~Abdikamalov, D.~Ayzenberg, C.~Bambi, J.~Jiang, H.~Liu and M.~Zhou,
{\it Testing General Relativity with NuSTAR data of Galactic Black Holes},
Astrophys. J. \textbf{913}, 79 (2021)
[arXiv:2012.10669 [astro-ph.HE]].

\bibitem{Zhang:2020qbx}
Z.~Zhang, H.~Liu, A.~B.~Abdikamalov, D.~Ayzenberg, C.~Bambi and M.~Zhou,
{\it Probing the near-horizon region of Cygnus X-1 with Suzaku and NuSTAR},
Phys. Rev. D \textbf{103}, 024055 (2021)
[arXiv:2012.01112 [astro-ph.HE]].

\bibitem{zuobin}
Z.~Zhang, H.~Liu, A.~B.~Abdikamalov, D.~Ayzenberg, C.~Bambi and M.~Zhou,
{\it Testing the Kerr black hole hypothesis with GRS 1716\textendash{}249 by combining the continuum-fitting and the iron-line methods},
[arXiv:2106.03086 [astro-ph.HE]].

\bibitem{Zhang:2019ldz}
Y.~Zhang, A.~B.~Abdikamalov, D.~Ayzenberg, C.~Bambi and S.~Nampalliwar,
{\it Tests of the Kerr hypothesis with GRS 1915+105 using different RELXILL flavors},
Astrophys. J. \textbf{884}, 147 (2019)
[arXiv:1907.03084 [gr-qc]].

\bibitem{ashutosh}
A.~Tripathi {\it et al.}, in preparation.

\bibitem{Xu:2018lom}
Y.~Xu, S.~Nampalliwar, A.~B.~Abdikamalov, D.~Ayzenberg, C.~Bambi, T.~Dauser, J.~A.~Garcia and J.~Jiang,
{\it A Study of the Strong Gravity Region of the Black Hole in GS 1354\textendash{}645},
Astrophys. J. \textbf{865}, 134 (2018)
[arXiv:1807.10243 [gr-qc]].

\bibitem{Tripathi:2020qco}
A.~Tripathi, M.~Zhou, A.~B.~Abdikamalov, D.~Ayzenberg, C.~Bambi, L.~Gou, V.~Grinberg, H.~Liu and J.~F.~Steiner,
{\it Testing general relativity with the stellar-mass black hole in LMC X-1 using the continuum-fitting method},
Astrophys. J. \textbf{897}, 84 (2020)
[arXiv:2001.08391 [gr-qc]].

\bibitem{Cardenas-Avendano:2019zxd}
A.~Cardenas-Avendano, S.~Nampalliwar and N.~Yunes,
{\it Gravitational-wave versus X-ray tests of strong-field gravity},
Class. Quant. Grav. \textbf{37}, 135008 (2020)
[arXiv:1912.08062 [gr-qc]].

\bibitem{Tripathi:2019bya}
A.~Tripathi, J.~Yan, Y.~Yang, Y.~Yan, M.~Garnham, Y.~Yao, S.~Li, Z.~Ding, A.~B.~Abdikamalov and D.~Ayzenberg, \textit{et al.}
{\it Constraints on the Spacetime Metric around Seven \textquotedblleft{}Bare\textquotedblright{} AGNs Using X-Ray Reflection Spectroscopy},
Astrophys. J. \textbf{874}, 135 (2019)
[arXiv:1901.03064 [gr-qc]].

\bibitem{Tripathi:2018bbu}
A.~Tripathi, S.~Nampalliwar, A.~B.~Abdikamalov, D.~Ayzenberg, J.~Jiang and C.~Bambi,
{\it Testing the Kerr nature of the supermassive black hole in Ark 564},
Phys. Rev. D \textbf{98}, 023018 (2018)
[arXiv:1804.10380 [gr-qc]].

\bibitem{Liu:2020fpv}
H.~Liu, H.~Wang, A.~B.~Abdikamalov, D.~Ayzenberg and C.~Bambi,
{\it Reflection features in the X-ray spectrum of Fairall 9 and implications for tests of general relativity},
Astrophys. J. \textbf{896}, 160 (2020)
[arXiv:2004.11542 [gr-qc]].

\bibitem{Choudhury:2018zmf}
K.~Choudhury, S.~Nampalliwar, A.~B.~Abdikamalov, D.~Ayzenberg, C.~Bambi, T.~Dauser and J.~A.~Garcia,
{\it Testing the Kerr metric with X-ray Reflection Spectroscopy of Mrk 335 Suzaku data},
Astrophys. J. \textbf{879}, 80 (2019)
[arXiv:1809.06669 [gr-qc]].

\bibitem{Nandra:2013jka}
K.~Nandra, D.~Barret, X.~Barcons, A.~Fabian, J.~W.~d.~Herder, L.~Piro, M.~Watson, C.~Adami, J.~Aird and J.~M.~Afonso, \textit{et al.}
{\it The Hot and Energetic Universe: A White Paper presenting the science theme motivating the Athena+ mission},
[arXiv:1306.2307 [astro-ph.HE]].

\bibitem{Zhang:2016ach}
S.~N.~Zhang \textit{et al.} [eXTP],
{\it eXTP -- enhanced X-ray Timing and Polarimetry Mission},
Proc. SPIE Int. Soc. Opt. Eng. \textbf{9905}, 99051Q (2016)
[arXiv:1607.08823 [astro-ph.IM]].

\bibitem{Riaz:2020zqb}
S.~Riaz, M.~L.~Szanecki, A.~Nied\'zwiecki, D.~Ayzenberg and C.~Bambi,
{\it Impact of the returning radiation on the analysis of the reflection spectra of black holes},
Astrophys. J. \textbf{910}, 49 (2021)
[arXiv:2006.15838 [astro-ph.HE]].

\end{thebibliography}
\end{document}